# Quantitative Analysis of Sodium Metal Deposition and Interphase in Na Metal Batteries

Baharak Sayahpour[1], Weikang Li[2], Shuang Bai[1], Bingyu Lu[2], Bing Han[2], Yu-Ting Chen[1], Grayson Deysher[1], Saurabh Parab[1], Phillip Ridley[2], Ganesh Raghavendran[2], Long Hoang Bao Nguyen[2], Minghao Zhang[2,*], Ying Shirley Meng[2,3,*]

[1]Materials Science and Engineering program, University of California San Diego, CA 92093, USA

[2]Department of NanoEngineering, University of California San Diego, CA 92093, USA

[3]Pritzker School of Molecular Engineering, University of Chicago, Chicago, IL 60637, USA

[*]Corresponding authors, miz016@eng.ucsd.edu, shirleymeng@uchicago.edu

## Abstract

Sodium-ion batteries exhibit significant promise as a viable alternative to current lithium-ion technologies owing to their sustainability, low cost per energy density, reliability, and safety. Despite recent advancements in cathode materials for this category of energy storage systems, the primary challenge in realizing practical applications of sodium-ion systems is the absence of an anode system with high energy density and durability. Although Na metal is the ultimate anode that can facilitate high-energy sodium-ion batteries, its use remains limited due to safety concerns and the high-capacity loss associated with the high reactivity of Na metal. In this study, titration gas chromatography is employed to accurately quantify the sodium inventory loss in ether- and carbonate-based electrolytes. Uniaxial pressure is developed as a powerful tool to control the deposition of sodium metal with dense morphology, thereby enabling high initial coulombic efficiencies. In ether-based electrolytes, the Na metal surface exhibits the presence of a uniform solid electrolyte interphase layer, primarily characterized by favorable inorganic chemical components with close-packed structures. The full cell, utilizing a controlled electroplated sodium metal in ether-based electrolyte, provides capacity retention of 91.84% after 500 cycles at 2C current rate and delivers 86 mAh/g discharge capacity at 45C current rate, suggesting the potential to enable Na metal in the next generation of sodium-ion technologies with specifications close to practical requirements.

## Introduction

The surge in global energy consumption and escalating environmental degradation has expedited the need for clean energy storage systems. Among the prevailing technologies, lithium-based energy storage systems, specifically lithium-ion batteries (LIBs), are recognized as pioneering solutions, finding extensive applications across various domains, ranging from compact portable electronics to electric vehicles.[1] The substantial demand for lithium resources and the concomitant rise in costs underscore the imperative to explore alternative or complementary clean technologies to mitigate reliance on lithium-based systems.[2,3] During the past decade, sodium-ion batteries (SIBs) have shown great promise for sustainable and cost-effective energy storage systems in grid-scale and transportation applications.[4–8] Despite advances in cathode materials for sodium-ion systems[9–11], development of a stable anode and electrolyte is still the key limiting factor in large-scale utilization of this battery technology.[12–14] In contrast to LIBs, the use of graphite as an anode material for SIBs is precluded by its inability to intercalate $Na^+$ ions within its structure.[15,16] Only $Na^+$ ions in the solvated state with diglyme (bis(2- methoxyethyl) ether) can be co-intercalated into graphite; however, the capacity of the reaction is too low that it can hardly find any practical applications.[17,18] Hard carbon was then introduced as a carbon-based negative electrode alternative to graphite; however, hard carbon is not an impeccable anode for SIBs due to several drawbacks: (i) relatively low specific capacity (~ 300 mAh/g), (ii) high irreversible capacity loss due to Na trapping in the first sodiation, and (iii) poor understanding of the sodiation mechanism.[19–21]

Among a wide range of possible anode materials for SIBs, sodium metal is the ultimate one thanks to its high theoretical specific capacity (1166 mAh/g) and low reduction potential (-2.73 V vs. standard hydrogen electrode) (**Table S1**).[22] Nonetheless, there exist several challenges hindering the large-scale utilization of sodium metal as a practical anode, mainly associated to its high chemical and electrochemical reactivity. Consequently, a substantial quantity of electrolyte is consumed during the initial cycle, giving rise to a pronounced formation process of the solid electrolyte interphase (SEI), thereby resulting in low initial coulombic efficiency (ICE) and diminished cyclability.[23–25] To overcome these challenges, several strategies have been proposed: (i) current collector modification by employing a porous three-dimensional structuring or an artificial coating that helps to enable a uniform electric field and ion-flux distribution on the electrode surface to lower the nucleation overpotential[26–32], (ii) electrolyte engineering to enable a stable and robust SEI layer for a uniform sodium nucleation[33–38], and (iii) application of an artificial SEI layer to mitigate direct contact between the metal electrode and electrolyte while enhancing the stability of the metal-electrolyte interface[39–43]. Regardless of the specific approach employed, the pivotal factor for enabling a sodium anode lies in the synergistic interplay among the electrolyte, SEI, and the sodium metal itself.

The application of stack pressure is an alternative approach that has long been discussed within the realm of lithium metal systems. Moly Energy Limited published a patent in 1985 on methods for making a battery that mentioned the stack pressure as a factor to control the preferable Li deposition.[44] Later, Hirai et al. showed that Li dendrite formation in Li metal anode can be controlled by applying uniaxial pressure using a coin-cell setup which leads to improve ICE and cycle life.[45] Recently, our group established the concept of "Pressure−Morphology−Performance" correlation as a rational design for improving the performance of Li metal batteries.[46–49] Similar behavior is expected to be observed for all metal anodes; nonetheless, to date, no comprehensive study has been documented regarding the Na metal anode. The exploration into the influence of applied pressure holds the potential to unveil a novel avenue for regulating the performance of Na metal, thereby enabling the realization of SIBs with enhanced energy density.

This study provides a comprehensive investigation into the impact of pressure on the morphology, electrochemical performance, and capacity loss mechanisms of Na metal anode in both ether-based and carbonate-based electrolytes. Notably, sodium inventory loss during the initial cycle at various applied pressures is evaluated for the first-time using titration gas chromatography (TGC). The findings highlight the primary role of SEI formation as the leading cause of Na metal inventory loss, which can be effectively mitigated through the application of an appropriate pressure on the cell. Through X-ray photoelectron spectroscopy (XPS) and cryogenic scanning transmission electron microscopy (cryo-STEM), a dense and uniform SEI layer with a dominant presence of organic species on the surface and inorganic species underneath is detected in ether-based electrolyte. On the other hand, the SEI layer in carbonate-based electrolytes is rather thick with a fluffy structure consisting of organic carbonyl and carboxyl species. Finally, a long-term cycling performance of a Na||$NaCrO_2$ cell is demonstrated using ether-based electrolytes, wherein a controlled amount of Na metal is employed. The Na metal is designed to exceed the Na inventory in the cathode active material by 100%. The cathode mass loading in this investigation is maintained at an average of 13 $mg/cm^2$. Notably, the cell exhibits a capacity retention of 91.84% after 500 cycles at the current rate of 2C. Additionally, the electrochemical performance of the system is evaluated under higher current rates, up to 45C, as well as at an elevated temperature of 40°C. The accomplishment documented in this study has the potential to pave the way for advancements in the development of high-energy SIBs through the utilization of a Na metal anode.

## Material and Methods

***Synthesis of $NaCrO_2$ Cathode Materials:*** Pure O3-phase $NaCrO_2$ (NCO) was synthesized from a stoichiometric ratio of $Cr_2O_3$ (99.97%, Alfa Aesar) and $Na_2CO_3$ (99.5%, Alfa Aesar). The mixture was pelletized and then calcinated under Ar at 900 °C for 10 h before being naturally cooled to room temperature.[50]

***Titration Gas Chromatography (TGC):*** TGC was performed using a Shimadzu GC instrument equipped with a BID detector and ultra-high purity Helium (99.999%) as the carrier gas. The samples were prepared in an Ar-filled glovebox with < 0.1 ppm $H_2O$ level. Each sample was immediately transferred to a glass flask after disassembling and sealed using a septum under Ar. A 0.5 mL of ethanol was injected into the container to fully react with metallic sodium. After reaction completion, a 30 μL gas sample was taken from the container using a gastight Hamilton syringe and immediately injected into the GC. The amount of metallic sodium was quantified based on the amount of detected $H_2$ gas by the GC. More details on TGC measurements are included in **Section S2**.

***Cryogenic Focused Ion Beam Scanning Electron Microscopy (Cryo-FIB-SEM):*** The FIB-SEM was conducted on the FEI Scios Dual-beam microscopy; the discharged cells were disassembled in the Ar-filled glovebox after cycling. The samples were transferred to the FIB chamber via quick loader without any exposure to air. The electron beam operating voltage was 5 kV, and the stage was cooled with liquid nitrogen to -180 °C or below. Sample cross-sections were exposed using a 1 nA ion beam current and cleaned at 0.1 nA. More details on cryo-FIB-SEM are included in **Section S4**.

***X-ray Photoelectron Spectroscopy (XPS):*** XPS was performed using an AXIS Supra by Kratos Analytica. XPS electrode samples were prepared inside an Ar-filled glovebox with < 0.1 ppm $H_2O$ level. Unwashed samples were directly dried under vacuum before measurements. The XPS was operated using an Al anode source at 15 kV, scanning with a step size of 0.1 eV and 200 ms dwell time. The etching condition used was $Ar^+$ mono mode, 5 keV voltage. The etching intervals were 60 s. XPS spectra was analyzed with CasaXPS software to identify the chemical composition on the surface of the electrodes.

***Cryogenic Scanning Transmission Electron Microscopy (STEM) and Electron Energy Loss Spectroscopy (EELS):*** The sample was mounted to an airtight cooling holder from Melbuild to eliminate any contaminations to the Na metal-containing samples and transferred to the TEM column directly. HR-TEM results were obtained on ThermoFisher Talos X200 equipped with a Gatan Oneview camera operated at 200 kV with low dose capability. The image was acquired with minimum beam damage at spot size 6 with a dose rate of ~200 e $Å^{-2}$ $s^{-1}$. The STEM-EELS data was collected through UltraFast DualEELS Spectrum Imaging detector with an exposure time of 0.02 s, and the dispersion energy was 0.25 eV per channel.

***Electrochemical Tests:*** The electroplating was performed in a custom-made pressure setup (details in **Section S3** and **Figure S4**) with Aluminum foil (MTI Corp.) as the current collector, rolled sodium metal as the counter electrode (Sigma Aldrich, ≥99%), and Celgard 2325 as the separator. The electrolytes were made using battery-grade sodium hexafluorophosphate ($NaPF_6$) salt from STREM Chemicals dissolved in dimethoxyethane (DME) from Sigma Aldrich (anhydrous, 99.5%), and in 1:1 wt.% ratio ethylene carbonate (EC) and dimethyl carbonate (DMC) solvent from GOTION (battery-grade). The molar concentration of $NaPF_6$ was kept constant at 1 M for all electrolytes. The sodium was plated at a current rate of 0.5 $mA/cm^2$ for a total capacity of 1 $mA/cm^2$ and was stripped at the same current rate with a cut-off voltage of 1 V.

The electrochemical performance of electroplated sodium versus NCO cathode was tested using CR2032 coin cells. The electroplated sodium initially was prepared in our custom-made pressure setup and then used as anode in the coin-cell setup. Pure O3-type NCO powder was synthesized using a solid-state synthesis method that developed in our group and used as the active material in cathode electrode. The cathode electrodes were prepared by casting the slurry (80 wt% NCO, 10 wt% super C65 conductive agent, and 10 wt% polyvinylidene fluoride (PVDF) binder) on Al foil and then dried overnight at 80°C under vacuum. The NCO theoretical capacity was considered as 120 $mAh/cm^2$ in different C rates. A controlled amount of 55 μL of 1M $NaPF_6$ in DME was used as the electrolyte. Detailed summary information of the coin-cell testing specifications are presented in **Table S3**.

***Electrochemical Impedance Spectroscopy (EIS):*** EIS was performed with an applied AC potential of 10 mV in the frequency range of 1 MHz to 0.01 Hz, using a Solartron 1260 impedance analyzer. The EIS measurements for each case were performed on the same cell setup in the three steps. More details on EIS measurements are included in **Section S5**.

## Results and Discussion

**Impact of electrolyte compositions on ICE of Na metal anode.** The ongoing battery research extensively relies on the utilization of ether- and carbonate-based electrolytes. Some of the commonly applied carbonate solvents include propylene carbonate (PC), ethylene carbonate (EC), dimethyl carbonate (DMC), and fluoroethylene carbonate (FEC). While dimethoxyethane (DME), diethylene glycol dimethyl ether (DEGDME/diglyme), and tetraethylene glycol dimethyl ether (TEGDME) are widely used ethers.

As the above solvents possess different dielectric constant, viscosity, and chain length, the solvation energy and thus the reactivity of solvated $Na^+$ ions would be different. In order to investigate the impact of solvents and salts on the ICE of Na metal anode, Na plating and stripping was performed with different solvent and salt combinations (**Figure S1**). Aluminum foil was used as the current collector for the Na plating/stripping experiment with rolled Na metal as the counter electrode. All the experiments were performed in the coin-cell configuration with an internal pressure of about 150 kPa. The obtained results reveal that ether-based electrolytes can enable high ICE (~60−80%), regardless of the nature of the salt (**Figure S2**). Conversely, the ICE observed in carbonate electrolytes is relatively low, necessitating the incorporation of additives such as FEC to enhance the overall performance of the cell. In a recent study on Li metal system, our group has reported that the irreversible capacity loss in the first cycle is mainly related to the dead lithium metal due to the loss of contact with the current collector rather than SEI formation.[47] These two contributions can be deconvoluted using TGC technique, in which the amount of dead metal can be quantified by measuring the amount of hydrogen gas released in the reaction with a proton source, such as water or ethanol, $M^0 + H^+ \rightarrow M^+ + \frac{1}{2} H_2$ (g). When working with Na metal, the use of ethanol is essential to prevent the formation of HF, which could potentially react with the aluminum current collector and introduce errors during quantification. A detailed explanation of TGC method for sodium metal quantification is given in **Section S2** and **Figure S3**. The TGC quantification after one cycle of plating and stripping reveals that SEI formation is the main cause (~75−85%) of inventory loss for Na metal anode regardless of the solvents or salts (**Figure S2**), which is consistent with the results in previous studies.[51–53] This behavior is completely different to that of Li, which agrees well with the chemistry of Li and Na. In comparison to lithium, sodium exhibits greater chemical reactivity, leading to rapid reactions with the components of the electrolyte to form SEI. Our previous studies showed that the applied pressure can play an essential role on the ICE value of Li metal anode.[48] Two representative electrolytes for the ether- and carbonate-based families, e.g. 1M $NaPF_6$ in DME and 1M $NaPF_6$ in EC:DMC (1:1), are thus chosen to demonstrate the effects of stacking pressure on the performance of Na metal.

**Pressure effect on the morphology and reversibility of plated Na metal anode.** In order to investigate the impact of stacking pressure, Na plating and stripping under controlled uniaxial pressure were performed using our custom-made setup[49,54] (**Section S3**). After one cycle, the Al current collector was recovered for TGC measurement. The obtained results show a correlation between the applied uniaxial pressure with the ICE, in which an optimal pressure is required to maximize the ICE values (**Figures 1a−b**). A maximum ICE of 88.2% under an optimal pressure of 180 kPa is attained using 1M $NaPF_6$ in DME, while a maximum ICE of 86.1% under an optimal pressure of 250 kPa is achieved using 1M $NaPF_6$ in EC:DMC (1:1). Note that the increase of the uniaxial pressure above the optimal values does not help to improve the ICE. Interestingly, the optimal uniaxial pressure for ether-based electrolyte is lower than that of carbonate-based electrolyte, e.g., ~180 kPa vs. ~250 kPa (**Figures 1a−b**). Even at the optimal pressure, SEI formation is still the main cause of the Na inventory loss, which is the intrinsic property of Na metal due to its high reactivity.

The morphology of sodium metal deposited at the optimal pressure was then evaluated using cryo-FIB-SEM. The beam sensitivity has been extensively discussed as one of the main limiting parameters in using electron microscopy for lithium.[55–57] This extreme instability is more severe for sodium element with a lower melting point (97.7°C for sodium versus 180.5°C for lithium at ambient pressure) and weaker atomic bonding than lithium.[12,58] Initial assessments and precautions on the necessity of using cryogenic condition for sodium metal imaging are discussed in **Section S4** and **Figures S5−S6**.

The cryo-FIB-SEM image of Na metal deposited at 10 kPa shows a dense sodium metal deposition in ether-based electrolyte. However, it shows a porous structure with whisker shape sodium deposition in the carbonate-based electrolyte (**Figures S7**), which agrees with the sodium platting results without applied pressure.[59] The cross-sectional image of Na metal plated under 180 kPa pressure in ether-based electrolyte shows a dense packing with no voids or porosity (**Figure 1c** and **Figure S8**). Similar morphology is also observed for deposited sodium in carbonate-based electrolyte under 250 kPa pressure; however, the packing is less dense with the presence of small voids at certain areas (**Figure 1d**). The cross-sectional images of electrodeposited sodium on a larger scale (**Figure S9**), obtained through plasma FIB-SEM, align consistently with the images presented in **Figure 1**.

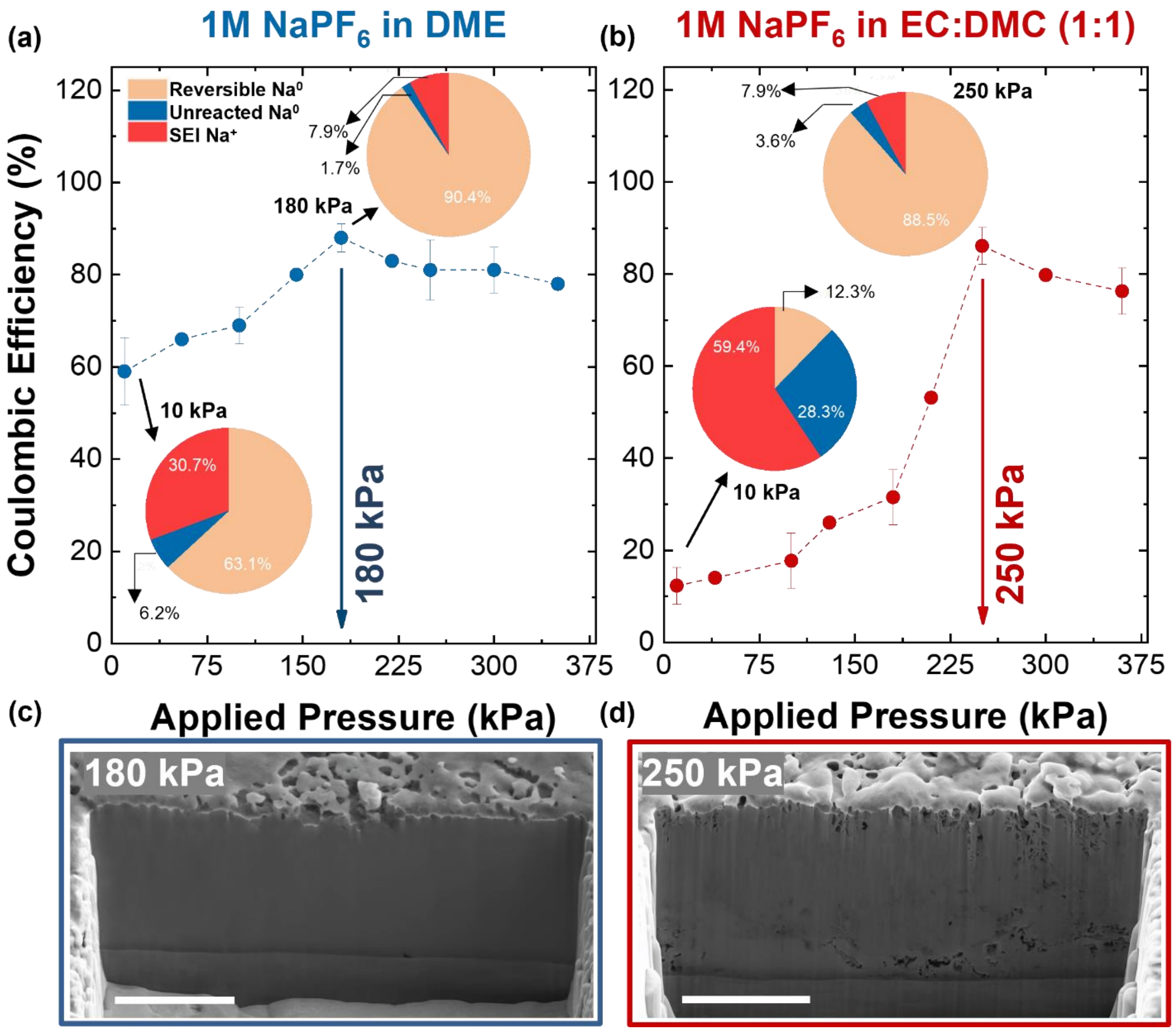

**Figure 1.** The first cycle CE of sodium on aluminum foil under different uniaxial pressure in **(a)** 1 M $NaPF_6$ in DME and **(b)** 1 M $NaPF_6$ in EC:DMC (1:1) electrolytes. Quantification of the Na inventory (reversible $Na^0$ metal, unreacted metallic $Na^0$, and $Na^+$ in the SEI) for minimum and optimal applied pressures using TGC method is also shown for each electrolyte. In these cells, sodium was plated at the current rate of 0.5 mA/cm$^2$ for a total capacity of 1 mA/cm$^2$ followed by stripping to 1 V at 0.5 mA/cm$^2$. Cross-sectional cryo-FIB-SEM images of the plated sodium at optimal pressures in **(c)** 1 M $NaPF_6$ in DME (180 kPa) and in **(d)** 1 M $NaPF_6$ in EC:DMC (1:1) (250 kPa). The scale bars are 5 μm.

Furthermore, thick sodium leftovers are visible after the half-stripping at 10 kPa in both electrolytes which agrees with the low ICE of the cell. At 250 kPa and 180 kPa, less sodium residues with uniform structure are observed after the half-stripping in carbonate- and ether-based electrolytes, respectively, in agreement with higher ICE of the cells (**Figure S10 and S11**). The porous network after stripping is a consequence of inhomogeneous SEI distribution in the plated Na layer. The presence of large voids and considerable residues at low pressures indicate a poor connection to Al current collector and further leads to the loss of the electronic conductive pathway. These observations indicate that applied uniaxial pressure greatly impacted the sodium nucleation and further facilitated the lateral growth on the surface of Al current collector.

EIS measurements were compared after the first platting and then the first stripping for both lowest (10 kPa) and optimal pressures (180 kPa for 1M $NaPF_6$ in DME and 250 kPa for 1M $NaPF_6$ in EC:DMC (1:1)) to obtain the cell resistance (**Section S5**). The Nyquist plots for the as assembled, after the $1^{st}$ plating, and after the $1^{st}$ stripping at 10 kPa and optimal pressures are shown in **Figure S15**. The sharp increase of the cell impedance (from as-assembled to plated to striped) at 10 kPa in EC:DMC should be mainly due to the thick porous sodium deposition that leads to a stronger barrier for the charge transfer and ion migration. And this resistance can be stabilized at the optimal uniaxial pressure of 250 kPa (**Figure S15f**). In contrast, a distinct decrease in cell impedance (from as-assembled to plated) in DME is in line with the dense morphology of plated sodium. Overall, the application of stack pressure results in lower interfacial resistance from ion migration through the SEI layer and charge transfer resistance for both electrolytes. The effect is more pronounced in the carbonate-based electrolyte. This observation can be attributed to disparities in both the morphology of plated sodium and the compositions of the SEI, which is discussed in the next section.

**Chemical composition of the SEI layer.** Cryo-STEM, cryo-EELS, and depth profiling XPS were applied to evaluate the SEI thickness and its chemical composition in both electrolytes. Depth profiling XPS was performed on the surface of the stripped sodium (after the first cycle) in both electrolytes, 1M $NaPF_6$ in DME and 1M $NaPF_6$ in EC:DMC (1:1) at the optimal pressures. The results at the core levels of carbon (C), sodium (Na), oxygen (O), fluorine (F), and phosphorus (P) are shown in **Figure 2**. The survey spectra presented in **Figure S16** show no other elements as contamination or impurities on the samples. The detection of the aluminum (Al) signal arises from the utilization of the current collector employed in the plating/stripping experiment. The atomic ratio of each element through the depth of etching on the SEI layer is summarized in **Figure 2a** for 1M $NaPF_6$ in DME and in **Figure 2b** for 1M $NaPF_6$ in EC:DMC (1:1). The data for the first step (outer surface) and the last step (after five steps of etching with 60s duration of etching in each step) are shown in these figures. The full dataset is provided in **Figure S17**.

Elemental evaluation in both electrolytes demonstrates a decrease in carbon content versus an increase in sodium content when moving away from the outer layer of the SEI. In carbonate-based electrolyte, the presence of carbon-containing species in the SEI is more dominant than in ether-based one (~30% vs. ~12% for the outer layer). More carbonyl and carboxyl species are detected in carbonate-based electrolyte. The presence of sodium species in the outer layer of SEI in carbonate-based electrolyte (~9%) is very limited compared to ether-based electrolyte (~21%). Na−O, $Na{-}CO_3$, Na−F, Na−PFO, and $Na{-}PO_4$ are Na species detected in both electrolytes; however, $Na{-}CO_3$ is dominant in carbonate-based electrolyte while Na−O, Na−F, Na−PFO, and $Na{-}PO_4$ share similar concentrations in ether-based electrolyte.

Fluorine and phosphorus are two other important elements in the chemical composition of SEI layers. $NaPF_6$ salt is the only source that can provide P and F in both electrolytes; however, the atomic ratio of these two elements is different when going through the depth of the SEI layer. Similar amounts of fluorine are detected in the inner layer of SEI in both electrolytes (~26% in ether- and ~29% in carbonate-based electrolyte). This content increases toward the outer layer of the SEI in ether-based electrolyte to 37% while decreases in carbonate-based electrolyte to 24%. Phosphorus content also shows higher content (~11%) closer to the outer layer of SEI in ether-based electrolyte with similar inner layer content (~5-6%) in both electrolytes.

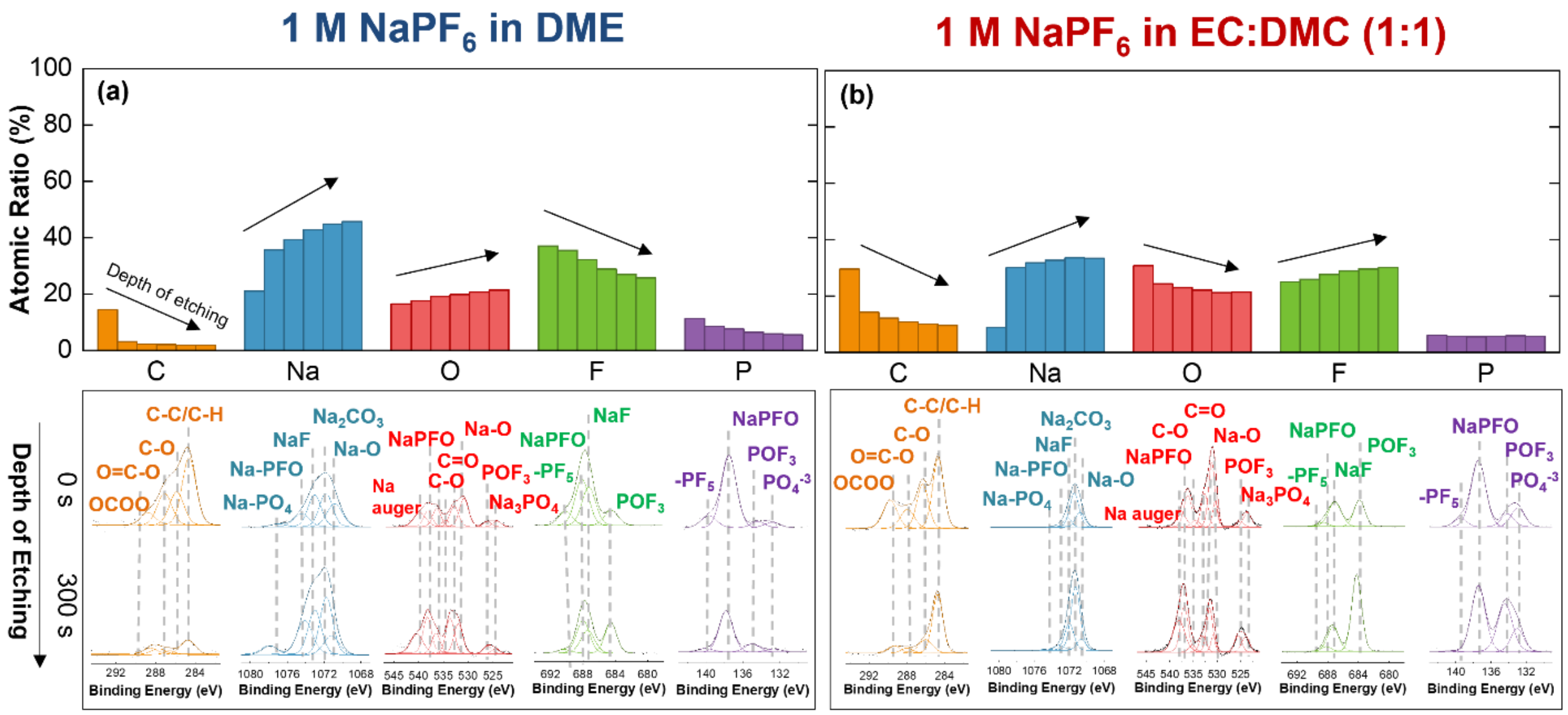

**Figure 2.** The characterization of the SEI layer using the depth profiling XPS on the stripped sodium sample in C 1s, Na 1s, O 1s, F 1s, and P 2p core levels. The atomic ratio of each element through the depth of etching (five steps of etching with 60s duration of etching) is presented for the case of **(a)** 1M $NaPF_6$ in DME, and **(b)** 1M $NaPF_6$ in EC:DMC (1:1). In these samples, sodium was initially plated at the rate of 0.5 mA/cm$^2$ for a total capacity of 1 mAh/cm$^2$, followed by a stripping to 1 V at the same rate.

The main species in SEI containing fluorine and phosphorus are Na−F, Na−$PO_xF_y$, and $POF_3$, and −$PF_5$. The Na−F and −$PF_5$ species mainly originate from the direct salt decomposition of $NaPF_6$ through the dissociation reaction.[60,61] On the other hand, Na−$PO_xF_y$, −$PO_4$ and $POF_3$ are largely resulted from the decomposition of carbonyl and carboxyl groups from the solvent and the hydrolysis reaction of the salt with water residues. Higher contribution of Na−F, Na−$PO_xF_y$, and −$PF_5$ in SEI layer of ether-based electrolyte confirms a more favorable salt participation. Moreover, despite the similar amounts of fluorine and oxygen in the inner layers of SEI in both electrolytes, the trend of the atomic ratio across the depth of SEI in these two elements is reversed. It is observed that fluorine increases while oxygen decreases toward the SEI outer layer in ether-based electrolyte and, in reverse; fluorine decreases and oxygen increases toward the SEI outer layer in carbonate-based electrolyte.

In order to further evaluate SEI structure and chemical compositions, cryo-STEM and cryo-EELS were also performed on the stripped sodium samples. The cryo-STEM images recorded on the stripped sodium under optimal pressures show a uniform and thin SEI layer (~25−30 nm) in DME (**Figures 3a**), which is completely different to a thick and fluffy SEI layer (~1500-2000 nm) in EC:DMC (**Figures 3b**). This observation is also in line with the difference in the interfacial and charge transfer resistances obtained from EIS measurements. The presence of a thicker SEI layer in carbonate-based electrolytes results in an elongated diffusion pathway, consequently impeding the electroplating process and leading to slower diffusion kinetics. The thickness and porosity of the SEI depends directly on the chemical composition of the layer as the components consisting of the SEI possess different stabilities, densities, and preferential growing directions of the lattice plane.

The polycrystalline nature of SEI components was then investigated using selected area diffraction (SAED) analysis in both electrolytes (**Figure 3c** and **Figure 3d**). The analysis reveals the presence of crystalline NaF, $Na_2CO_3$, and $Na_2O$ in the SEI structure of both electrolytes. In DME electrolyte, the SEI structure exhibits dominant phases of $Na_2O$. Furthermore, in EC:DMC electrolyte, $Na_3PO_4$ and Na-$PO_3$ are also identified, while Na-$PO_xF_y$ is detected in DME electrolyte. These observations are consistent with the previously proposed hypothesis of salt and/or solvent decomposition, as indicated by XPS results.

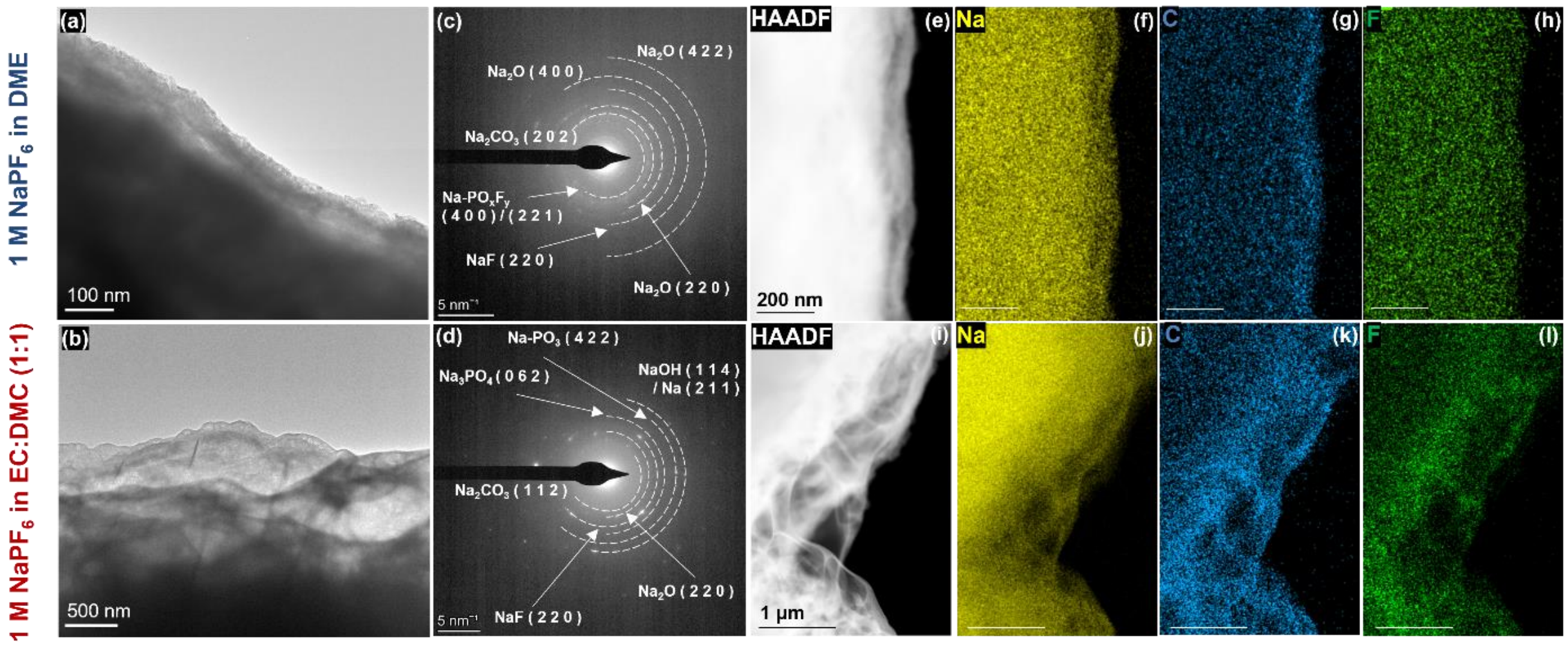

**Figure 3.** The cryo-STEM images on the SEI of the stripped sodium in **(a)** 1M $NaPF_6$ in DME, and **(b)** 1M $NaPF_6$ in EC:DMC (1:1). The SAED patterns show the crystalline structure of SEI components for **(c)** ether-based and **(d)** carbonate-based electrolytes. EDS elemental mappings are presented for **(e)** ether-based electrolyte in **(f)** Na, **(g)** C, and **(h)** F regions, and for **(i)** carbonate-based electrolyte in **(j)** Na, **(k)** C, and **(l)** F regions.

EDS elemental mapping was also conducted to investigate the chemical composition of SEI structures in ether-based (**Figure 3e-i** and **S18**) and carbonate-based (**Figure 3j-m** and **S19**) electrolytes. The mapping reveals a prevalence of carbon on the SEI outer layer, whereas sodium was observed to be more abundant in inner regions. These observations, together with the EELS chemical mappings (**Figure S20** and **S21**), are consistent with the depth profiling XPS results previously discussed. Cryo-EELS spectra of the C K-edge, Na K-edge, and O K-edge (**Figure S22**) are compared to those of the $Na_2CO_3$ reference sample, further confirming the presence of crystalline $Na_2CO_3$ in the SEI structures.

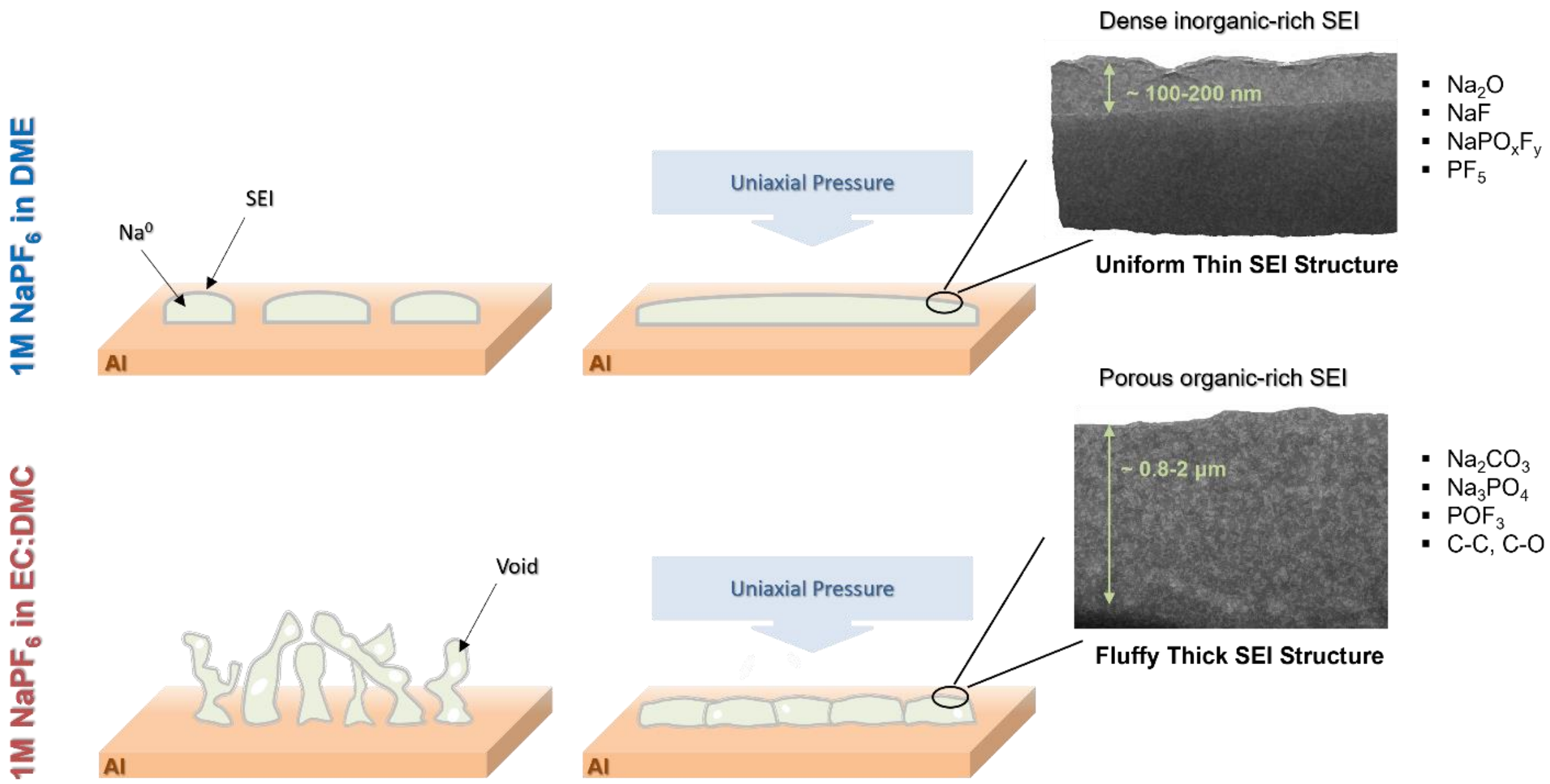

**Figure 4.** The schematic of the proposed sodium deposition and interphase under uniaxial pressure in ether and carbonate -based electrolytes.

A schematic of the proposed sodium deposition and interphase under uniaxial pressure in carbonate- and ether- based electrolytes is summarized in **Figure 4**. Under a uniaxial pressure of 10 kPa, the deposition of sodium metal in the ether-based electrolyte demonstrates a compact morphology characterized by islanded structures. In contrast, in the carbonate-based electrolyte, the deposited sodium exhibits a porous structure with whisker-shaped deposits. Increasing the uniaxial pressure can play an important role in electrochemical nucleation as it promotes the lateral sodium deposition and densifies the individual sodium particle through smoothing the surfaces. It is observed that sodium plated under the optimal pressure of 180 kPa in the ether-based electrolyte reveals a densely packed structure without any presence of voids or porosity. Moreover, in the carbonate-based electrolyte, the deposition of sodium under the optimal pressure of 250 kPa shows a dense packing, although certain regions exhibit small voids.

Furthermore, the SEI layer possesses more inorganic components (rich in sodium, fluorine, and oxygen) in ether-based electrolyte as opposed to more organic species (rich in carbon and oxygen) in carbonate-based electrolyte. In ether-based systems, a thin and dense SEI layer is observed, owing to the crystallization of inorganic SEI components such as NaF and $Na_2CO_3$ into close-packed structures. Conversely, carbonate-based electrolytes foster the formation of a fluffy and porous SEI layer, attributed to the presence of organic components like esters, which typically possess long organic chains that hinder the material from compact packing. This hypothesis is also supported by lower ionic conductivity of $Na_2CO_3$ ($5.69 \times 10^{-23}$ S/cm at 25°C) compared to $Na_2O$ ($1.47 \times 10^{-12}$ S/cm at 25°C) and easier deformation of $Na_2CO_3$ due to lower Young and Shear Moduli (31.9 and 11.47 GPa for $Na_2CO_3$ vs. 76.34 and 31.2 GPa for $Na_2O$).

**Electrochemical evaluation of Na metal full cell.** Uniaxial pressure has been shown as a powerful knob to control the morphology of plated Na in organic electrolytes. Optimal pressure of ~180 kPa is determined to yield a dense Na layer with a favorable composition and a uniform, thin SEI layer using 1M $NaPF_6$ in DME electrolyte. These results hold the potential to facilitate the development of long-term cycling batteries, offering precise control over the quantity of Na metal employed as the anode material. To demonstrate this concept, Na||1M $NaPF_6$ in DME||$NaCrO_2$ cells were assembled and tested at different C rates and higher temperature at 40°C. The Na anode was obtained through electroplating on a carbon-coated aluminum foil under the optimal pressure of 180 kPa and the current rate of 0.5 mA/cm$^2$ in DME electrolyte using our custom-made pressure setup. The amount of electroplated sodium is designed to be 100% excess to the Na inventory in cathode active material. The detailed calculations are provided in **Section S8**. The electroplated sodium was then used as anode in a coin-cell with NCO cathode and DME-based electrolyte. It should be noted that the applied pressure inside a coin-cell is about 150 kPa[49] that is around the optimal value for sodium deposition in the ether-based electrolyte. The electrochemical performance of the cell over 500 cycles (at 2C current rate) is presented in **Figure 5a** and **Figure S23**. The cell shows an average CE of 99.91% and capacity retention of 91.84% at room temperature. At 40°C, the cell shows an ICE of 95.08% at C/3 and an average CE of 99.68% with capacity retention of 96.29% after 100 cycles (**Figure S24**).

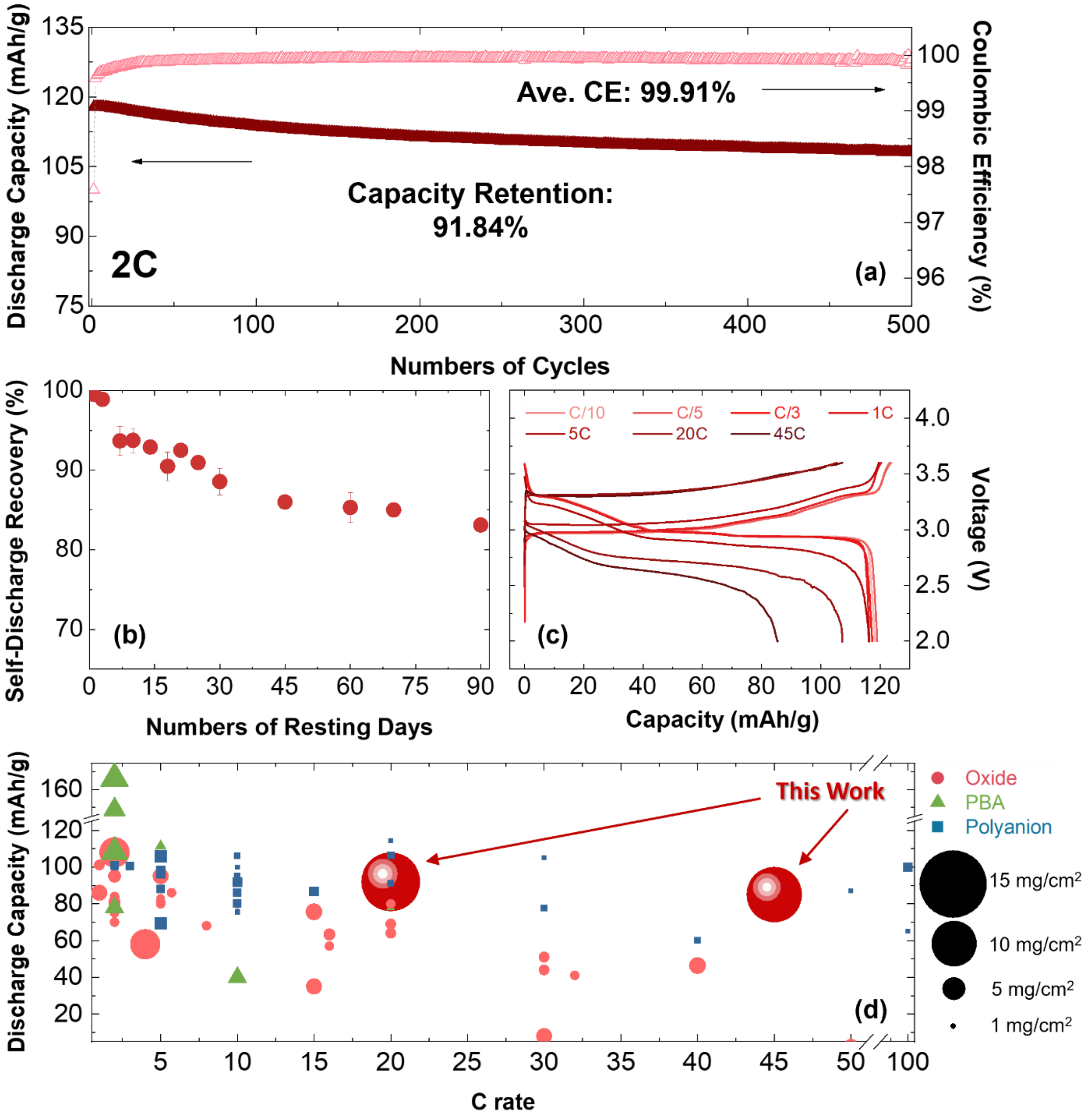

**Figure 5.** The electrochemical performance of the cell consisting of electroplated sodium as the anode, $NaCrO_2$ as the cathode, and 1M $NaPF_6$ in DME as the electrolyte. The cells have controlled 100% excess of sodium inventory. **(a)** Discharge capacity (mAh/g) and CE (%) versus numbers of cycles at the rate of 2C. **(b)** The rate of self-discharge recovery at room-temperature. **(c)** The first cycle voltage profiles of cell at different C rates of C/10, C/5, C/3, 1C, 5C, 20C, and 45C. **(d)** The standing of this work at varied high rates (dark red circles) compared to the literature with various cathode structures (red circle: oxide, green triangle: PBA, blue square: polyanion) and different loadings (from 1 to 15 $mg/cm^2$).

Another critical aspect in assessing the performance of metal batteries is the self-discharge rate during a period of rest. When sodium metal comes into direct contact with the liquid electrolyte, a spontaneous chemical corrosion process can occur, leading to detrimental effects on battery performance. This chemical reaction between sodium and the electrolyte gradually consumes the active sodium metal, resulting in increased cell impedance, reduced shelf life, and ultimately, cell failure. The formation of a passivating layer known as the SEI on the surface of sodium metal can mitigate these adverse effects. The uniformity of the formed SEI layer, as well as the ionic and electronic conductivity of its components, play a pivotal role in controlling the chemical corrosion process and effectively lowering the self-discharge rate.

To shed light on the self-discharge rate of sodium metal in this study, the self-discharge characteristics of the system were assessed according to the procedures outlined in **Section S9** and depicted in **Figure S26**. The self-discharge rate was determined from the recovery rate, as shown in **Figure 5b**. The results indicate a rapid decrease in capacity during the initial 15−30 days of the test, followed by a slower decline over the subsequent 50 days, suggesting an initial rapid formation of SEI that becomes more stable. The self-discharge rate is approximately 12% after the first 30 days of testing, and this value increased to 17% after 90 days.

Furthermore, the electrochemical performance of the system was evaluated at different current rates (at C/10, C/5, C/3, 1C, 5C, 20C, and 45C) and the voltage profiles are presented in **Figure 5c**. The obtained results indicate that the cells can operate at high current rates, delivering 107 mAh/g and 86 mAh/g discharge capacities at current rates of 20C and 45C, respectively. It is worth noting that achieving high current performance has still remained a persistent hurdle in lithium-ion batteries where fast charging leads to accelerated performance degradation and reduced energy efficiency, primarily attributed to issues such as lithium diffusion in the lattice structure of cathode material, lithium plating on the anode side, and overall heat generation.[62,63] Subsequently, this performance degradation becomes evident through the emergence of higher voltage polarization and capacity loss. Notably, unlike lithium-ion chemistry, the findings of this study emphasize a similar high-rate compatibility in both charging and discharging reactions in sodium metal batteries. This effect is attributed to (i) a high thermal and structural stability of $NaCrO_2$ cathode material, and (ii) a stable, uniform, and thin SEI layer on the sodium anode promoting reaction kinetics. Recent literatures have brought attention to the robust structural integrity demonstrated by the $NaCrO_2$ cathode material in sodium ion batteries when subjected to elevated charge/discharge rates.[64,65] However, utilization of a high-loading $NaCrO_2$ cathode along with carbonate-based electrolyte resulted in a charge/discharge rate below 10C (**Figure S25**), indicating that the critical parameter is not solely determined by the cathode. Furthermore, the importance of SEI characteristics on the reaction kinetics has been demonstrated in recent studies.[67–69] This encloses the physicochemical features of SEI layer, such as morphology and compositional properties, that reveals to play a pivotal role in mitigating electron leakage and self-discharge phenomena. This further emphasizes our approach in implementing a pressure-controlled setup to achieve a uniform and thin inorganic SEI layer. In this context, as previously discussed, the Nyquist plots obtained from EIS measurements on the Na||Al system under pressure-controlled conditions utilizing ether-based and carbonate-based electrolytes are detailed in **Section S5**. The collective findings indicate that the contribution of bulk electrolyte diffusion resistance ($R_b$) remains negligible, accounting for less than 1%, in stark contrast to the substantial contribution of charge transfer resistance and interfacial resistance from the ion migration through the SEI layer ($R_1$+$R_2$), exceeding 99%. In addition, the ion migration through the SEI layer and charge transfer resistance exhibits an augmentation in both electrolyte systems after the stripping process. In the case of the ether-based electrolyte, this resistance experiences a 30% escalation post-stripping, while in the carbonate-based electrolyte, a more pronounced increase of approximately 73% was observed compared to that after plating process. Overall, the precise manipulation of Na deposition and stripping through pressure control and electrolyte engineering as well as the robust structure of $NaCrO_2$ cathode are critical steps to enable fast charging operation for Na metal batteries. Sodium fast charging achieved in this study can provide several benefits for further applications, including higher energy density, improved power output, enhanced efficiency, reduced downtime, and cost-effectiveness.[70]

To further highlight the capabilities of the system, a comparison is made with relevant literature on various cathode structures (oxides, PBAs, and polyanions) and cathode loadings, considering different applied current rates. **Figure 5d** depicts the comparison, demonstrating the robustness and superior performance of the system investigated in this study when subjected to higher current rates (up to 45C) and high cathode material loading (>10 mg/cm$^2$). It should be mentioned that sodium foil with no control on the amount of excess sodium metal is commonly used in the reported studies. The Na metal full cell with $NaCrO_2$ oxide as the cathode material in this study exhibits remarkable stability at a higher current rate of 45C, while other reported oxide cathode materials typically operate within the range of less than 10C. Additionally, reported polyanion-based cathode materials capable of performing at high rates (>10C) are only demonstrated at limited cathode loadings (<4 $mg/cm^2$) with Na metal anode. This lower cathode loading fails to meet the industrial requirements and restricts the practical application of these systems.

## Conclusions

In this study, the impact of uniaxial pressure on the growth of sodium metal in carbonate- and ether-based electrolytes is investigated. For each system, an optimal pressure is identified, leading to the highest ICE during the plating and stripping processes of sodium. This improved performance is enabled through the formation of a dense electroplated Na layer at the optimal pressure. Interestingly, irrespective of the solvent utilized, the optimal pressure required for sodium is considerably lower than that for lithium, potentially attributable to the comparatively lower Young's modulus of sodium metal. Additionally, the inventory loss experienced by the sodium metal anode is determined to be primarily attributed to the formation of the SEI, which aligns with the high reactivity exhibited by sodium metal. The SEI thickness and its chemical compositions have been shown to depend strongly on the type of electrolyte. Ether-based electrolyte enables a thin and dense SEI, while a fluffy and porous SEI is formed in carbonate-based electrolyte. In order to enable Na metal anode for practical applications, two essential parameters must be taken into account: (i) uniaxial pressure which controls the uniformity and thickness of the electroplated Na layer, and (ii) nature of the solvent and salt which has direct impact on the thickness and chemical compositions of the SEI layer. With the aforementioned approach, the performance of sodium metal batteries using a controlled amount of sodium metal anode is demonstrated. The system showcases a capacity retention of 91.84% after 500 cycles at 2C current rate. Furthermore, it exhibits an 86 mAh/g discharge capacity at a high rate of 45C. Such findings contribute significantly to the practical development of the next generation of sodium battery technologies.

## Acknowledgment

This work is sponsored in part by the UC San Diego Materials Research Science and Engineering Center (UCSD MRSEC), supported by the National Science Foundation (Grant DMR-2011924). This work was performed in part at the San Diego Nanotechnology Infrastructure (SDNI) of UCSD, a member of the National Nanotechnology Coordinated Infrastructure, which is supported by the National Science Foundation (Grant ECCS-1542148). The authors also acknowledge the use of facilities and instrumentation supported by NSF through the UC San Diego Materials Research Science and Engineering Center (UCSD MRSEC), DMR-2011924. The authors acknowledge the use of facilities and instrumentation at the UC Irvine Materials Research Institute (IMRI), which is supported in part by the National Science Foundation through the UC Irvine Materials Research Science and Engineering Center (DMR-2011967); specifically, the XPS work was performed using instrumentation funded in part by the National Science Foundation Major Research Instrumentation Program under grant No. CHE-1338173. The authors would like to acknowledge Neware for the generous donation of BTS4000 cyclers. The authors also thank Dr. Ich Tran at UCI IMRI facility and Dr. Mehdi Chouchane at the University of Chicago.

## Author Contributions

B.S., M.Z., and Y.S.M. conceived the idea. B.S. and M.Z. designed the experiments. B.S. performed and processed the data for TGC, EIS and the electrochemical performances. B.L.

performed the cryogenic FIB. Y.C. performed the plasma FIB. W.L. performed XPS experiment. B.S. processed the data for XPS. S.B. and B.H. performed the cryogenic STEM and EELS. G.D., P.R., and L.H.B.N. synthesized the pure NCO cathode material. All authors discussed the results and contributed to the manuscript. All authors have approved the final version of the manuscript.

**Declaration of Interests**

The authors declare no competing financial interest.